\documentclass{article}
\usepackage[utf8]{inputenc}
\usepackage[T1]{fontenc}
\pdfoutput=1

\usepackage{geometry}
\usepackage{cite}

\usepackage{authblk}

\usepackage{textcomp} 
\usepackage{gensymb}  
\usepackage{amssymb}  

\usepackage{array}
\newcolumntype{M}[1]{>{\centering\arraybackslash}m{#1}} 

\usepackage{graphicx}
\graphicspath{{./Images/}}
\usepackage[font=small, labelfont=bf]{caption}

\usepackage{balance}

\usepackage{hyperref}
\title{Optimization of MR Fingerprinting for Free-Breathing Quantitative Abdominal Imaging}
\author[1,2,3]{Max H.C. van Riel}
\author[2,3,4]{Zidan Yu}
\author[2,3,4,5]{Shota Hodono}
\author[2,3]{Ding Xia}
\author[2,3]{Hersh Chandarana}
\author[6,7]{Koji Fujimoto}
\author[2,3,4,5]{Martijn A. Cloos}
\affil[1]{Department of Biomedical Engineering, Eindhoven University of Technology, Eindhoven, The Netherlands}
\affil[2]{Bernard and Irene Schwartz Center for Biomedical Imaging, Department of Radiology, New York University School of Medicine, New York, NY, USA}
\affil[3]{Center for Advanced Imaging Innovation and Research (CAI\textsuperscript{2}R), Department of Radiology, New York University School of Medicine, New York, NY, USA}
\affil[4]{Sackler Institute of Graduate Biomedical Sciences, New York University School of Medicine, New York, NY, USA}
\affil[5]{Centre for Advanced Imaging, University of Queensland, Brisbane, Queensland, Australia}
\affil[6]{Human Brain Research Center, Graduate School of Medicine, Kyoto University, Kyoto, Japan}
\affil[7]{Department of Real World Data Research and Development, Graduate School of Medicine, Kyoto University, Kyoto, Japan}
\date{}

\begin{document}

\maketitle

\begin{abstract}
    \noindent \textbf{Purpose:} To obtain $B_1^+$-robust quantitative maps of the abdomen using free-breathing magnetic resonance fingerprinting in a clinically acceptable time. 
\\
\\
\textbf{Methods:} A three-dimensional MR fingerprinting sequence with a radial stack-of-stars trajectory was implemented for quantitative abdominal imaging. The flip angle pattern was optimized using the Cramér-Rao Lower Bound, and the encoding efficiency of sequences with 300, 600, 900, and 1800 flip angles was evaluated. To validate the sequence, a movable multicompartment phantom was developed. Reference multiparametric maps were acquired under stationary conditions using a previously validated MRF method. Periodic motion of the phantom was used to investigate the motion-robustness of the proposed sequence. The best performing sequence length (600 flip angles) was used to image the abdomen during a free-breathing volunteer scan.
\\
\\
\textbf{Results:} When using a series of 600 or more flip angles, the estimated $T_1$ values in the stationary phantom showed good agreement with the reference scan. Phantom experiments revealed motion-related artefacts can appear in the quantitative maps, and confirmed that a motion-robust k-space ordering is essential in preventing these artefacts. The in vivo scan demonstrated that the proposed sequence can produce clean parameter maps while the subject breathes freely.
\\
\\
\textbf{Conclusion:} Using this sequence, it is possible to generate $B_1^+$-robust quantitative maps of proton density, $T_1$, and $B_1^+$ under free-breathing conditions at a clinically usable resolution within 5 minutes.

\end{abstract}


\twocolumn

\section{Introduction}
Most routine clinical Magnetic Resonance Imaging (MRI) measurements depict a relative contrast between tissues. In addition to the tissue properties, the measured signal intensities depend on experimental factors, such as RF excitation and receive sensitivities \cite{Brown2014}. Consequently, the obtained qualitative contrast depends on the specific sequence parameters and scanner hardware used to acquire the data. Quantitative MRI, on the other hand, strives to directly measure physical or chemical properties of tissues, most notably the longitudinal ($T_1$) and transverse ($T_2$) relaxation times. Compared to qualitative images, such quantitative maps enable more straightforward comparisons between scans from different patients, different timepoints, or using different hardware \cite{Cheng2012, Deoni2010}. 

Quantification of MR parameters can be done using several techniques. Historically, relaxometry measurements were performed by fitting exponentials to a series of inversion times or echo times to quantify $T_1$ or $T_2$ relaxation times, respectively \cite{Brown2014}. However, such measurements are too slow for routine clinical usage. Over the years many techniques have been developed to provide a better balance between accuracy and acquisition speed, such as the Look-Locker method \cite{Look1970} or DESPOT1 and DESPOT2 \cite{Deoni2003}, among others. Recently, MR fingerprinting (MRF) has been proposed for the fast and simultaneous quantification of multiple parameters \cite{Ma2013}. This method uses variable flip angle train that produces unique signal evolutions, so-called ``fingerprints'', for different tissue parameter combinations. By comparing the measured fingerprints with the entries in a precomputed dictionary, the underlying tissue parameters, such as $T_1$, $T_2$, and proton density (PD), can be estimated.

Several challenges arise when imaging the abdomen. Intestinal gas in the abdomen can cause susceptibility artefacts, especially for field strengths of 3T and higher \cite{Merkle2006}. Furthermore, the relatively short wavelength of the RF can lead to interferences in the $B_1^+$ field, resulting in an inhomogeneous excitation \cite{Merkle2006, Collins2005}. Moreover, respiratory, cardiac, gastrointestinal, and voluntary motion can cause other image artefacts \cite{Yang2010} and can corrupt the MRF-based parameter maps \cite{Yu2018}. Although breath-holds can be used to eliminate respiratory motion during an MRF experiment \cite{Chen2016}, it places restrictive limits on scan time and thus on spatial resolution or volumetric coverage. Besides, not all patients can hold their breath (e.g. pediatric patients) \cite{Bernstein2004HandbookSequences, Masaracchia2017}. Therefore, it is preferable that the sequence is inherently robust against motion, such that the subject can breathe freely during the scan \cite{Block2014}. 

In this work, we present a free-breathing 3D MRF sequence to generate quantitative PD, $T_1$, and $B_1^+$ maps. The k-space ordering was adjusted to improve robustness against motion, while $B_1^+$-related artefacts were addressed by including $B_1^+$ in the parameter estimation. The flip angles were optimized to improve the efficiency of the sequence and to decrease the scan time. The method was validated using a movable phantom and subsequently evaluated in vivo during free-breathing abdominal measurements.

\section{Methods}
\subsection{Sequence Design}
Building on prior work \cite{Cloos2019}, an MRF sequence was designed to generate a distinct signal pattern for every combination of $T_1$ and $B_1^+$. By including the $B_1^+$ value in the parameter estimation, the effect of an inhomogeneous excitation field can be dealt with \cite{Cloos2016, Buonincontri2017}. Since most $B_0$-robust $T_2$ encoding solutions are particularly motion sensitive \cite{Yu2018}, only fast low-angle shot (FLASH) segments with a repetition time (TR) of 5 ms were used. These segments were gradient- and RF-spoiled to avoid $T_2$ dependence.

To reduce the sensitivity of the sequence to motion, a radial readout sampling pattern was chosen \cite{Block2014}. This trajectory samples the center of k-space, which encodes the global image contrast, with every readout. Consequently, the motion during acquisition is averaged out in the resulting image. Furthermore, undersampling artefacts are expressed as streaks in the image \cite{Scheffler1998}, instead of the less desirable ghosts that occur with a Cartesian sampling pattern. These streaks lead to incoherent, noise-like artefacts in the reconstructed images, which can be filtered out by the dictionary matching process. The extension to 3D can be made by stacking multiple partitions on top of each other, where each partition consists of several radial readouts, to form a stack-of-stars trajectory \cite{Block2014}. 

\begin{figure*}[!t]
    \centering
    \includegraphics[width=12cm]{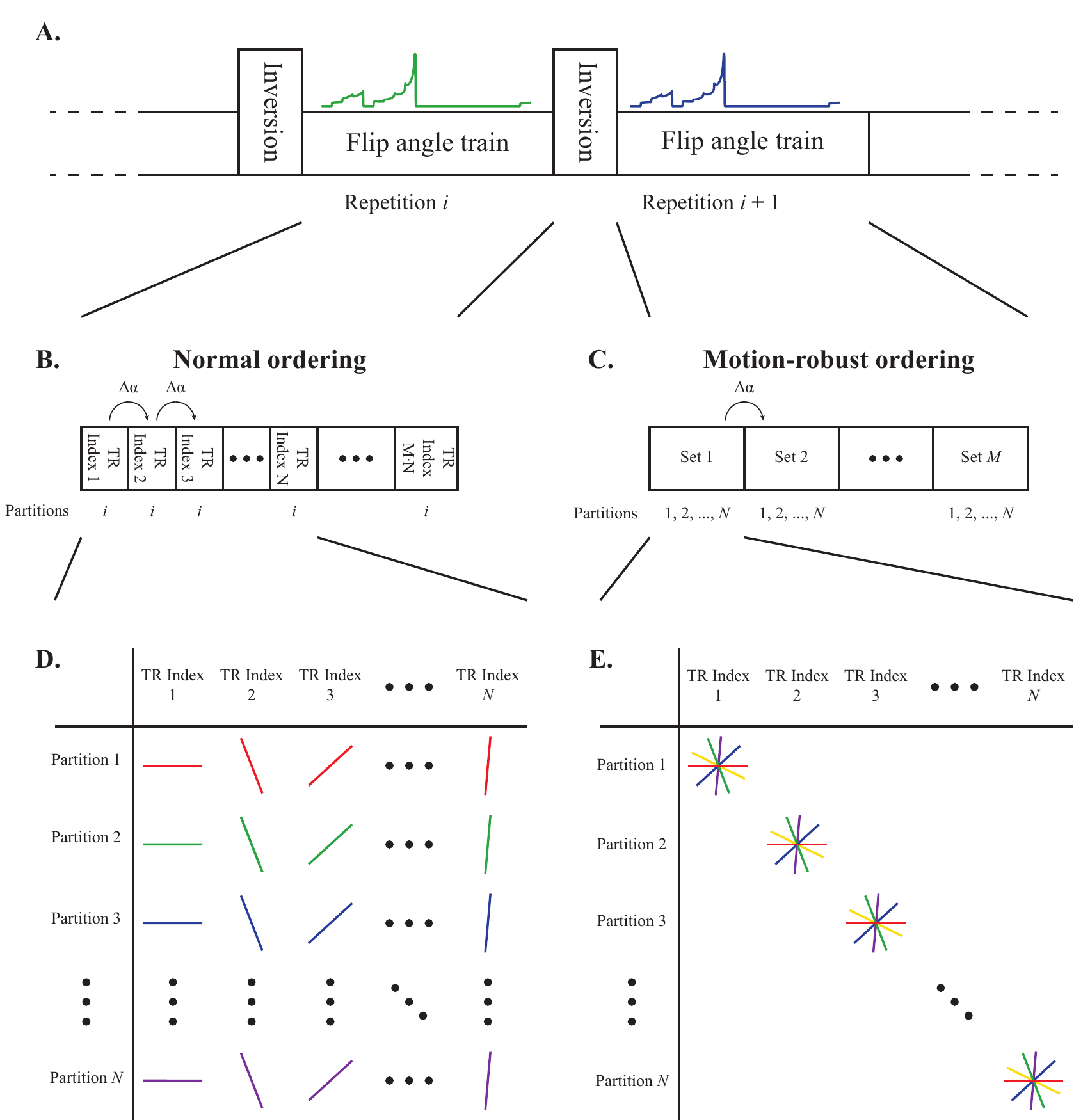}
    \caption{Proposed motion-robust 3D MRF sequence. \textbf{A}: The data is acquired using a repeated flip angle sequence, each generating a fingerprint encoding $T_1$ and $B_1^+$. Every repetition is preceded by a non-selective adiabatic inversion pulse. \textbf{B}: In the \textit{normal ordering}, each fingerprint sequence consists of $M \times N$ TRs. Each TR index generates a separate k-space. \textbf{C}: In the \textit{motion-robust ordering}, each fingerprint sequence consists of $M$ sets, containing $N$ TRs each. All readouts acquired during the TR indices corresponding to one set are grouped together to form a single k-space. \textbf{D}: Readout lines acquired for the \textit{normal ordering} during the first $N$ TRs. Every TR, the readout angle is incremented with the golden angle ($\Delta \alpha = 111.25\degree$), while the partition index stays the same. This results in a single undersampled k-space for every TR index. All lines of the same color are acquired sequentially during the same repetition. \textbf{E}: Readout lines acquired for the \textit{motion-robust ordering} during the first set. Every TR, the readout angle stays the same, but the partition index is increased. Note how readouts in adjacent partitions with the same angle are acquired successively in the \textit{motion-robust ordering}, while in the \textit{normal ordering}, the next partition is sampled during the next repetition of the sequence.}
    \label{fig:SequenceDesign}
\end{figure*}

The simplest stack-of-stars implementation, hereafter referred to as \textit{normal ordering}, repeats the flip angle train for every partition, while increasing the readout angle by 111.25 degrees with every readout \cite{Winkelmann2007}. This way, adjacent lines in the phase encoding direction (which is along the partitions) are acquired using the same flip angle, and thus have the same contrast weighting (Figure \ref{fig:SequenceDesign}). However, this ordering scheme results in a time interval of several seconds between adjacent lines in the partition direction, and any motion during this time results in motion artefacts in the fingerprints. 

To obtain motion robustness while using a stack-of-stars sampling pattern, all partitions are typically acquired in quick succession before changing the readout angle \cite{Block2014}. However, during an MRF experiment, the MR signal is in a transient state. Consequently, signals obtained with different flip angles are mixed during the Fourier transform along the phase encoding direction. To allow reconstruction of the data, the measured signals are divided into sets, with each set containing the data from as many consecutive TRs as the number of partitions (hereafter referred to as \textit{motion-robust ordering}, see Figure \ref{fig:SequenceDesign}). When the flip angles change smoothly, it is our hypothesis that the differences in signal intensities along the partition direction will remain acceptable.

The flip angle train was repeated multiple times. Data acquisition was started from the second repetition onwards to ensure that the initial magnetization at the start of each repetition was identical. Next, the sequence was repeated once for every partition (each time with the same readout angle in k-space), and once for every shot (with different readout angles in k-space). The number of shots is defined as the number of unique readout angles per TR in the flip angle train. The total number of repetitions is thus equal to $N_{shots} \times N_{partitions}+1$.

\subsection{Dictionary Construction}
The Bloch equations \cite{Bloch1946} were used to create a precomputed dictionary consisting of 17600 simulated fingerprints, each with a unique combination of $T_1$ and $B_1^+$ values. The $T_1$ values ranged from 50 ms to 3764 ms with relative increments of 2.5\%, and $B_1^+$ effects were implemented by multiplying the desired flip angle with a relative weighting factor ranging from 0.02 to 2.0 in steps of 0.02. Simulating $T_2$ effects was not necessary since the combination of gradient- and RF-spoiling eliminates all stimulated echoes. In addition, the echo time (TE) was held fixed, resulting in a constant scaling of the fingerprints caused by $T_2^*$ effects. 

For every combination of $T_1$ and $B_1^+$ in the dictionary, the whole sequence was simulated for two repetitions. The first repetition was used to obtain the initial magnetization during a continuous scan, while the second repetition was used to obtain the fingerprint. Next, for the \textit{motion-robust ordering}, the fingerprint signals from $N_{partitions}$ consecutive readouts were grouped into sets (as in Figure \ref{fig:SequenceDesign}), and the signals within each set were averaged. The dictionary was compressed by projecting the fingerprints onto the first 5 singular vectors of the dictionary \cite{McGivney2014SVDDomain}. This rank-5 approximation retained more than 99.5\% of the energy in all dictionaries, where the energy is defined as the sum of the squared singular values. Finally, all simulated and averaged fingerprints were normalized to have unit Euclidean norm.

\subsection{Sequence Optimization}

Besides making the sequence motion robust, the flip angle pattern was optimized to bring the acquisition time down to under 5 minutes. Four different patterns were designed, corresponding to a sequence of 300, 600, 900, and 1800 flip angles. Consecutive flip angles were placed 5 ms apart, with a non-selective adiabatic inversion pulse \cite{Ordidge1996} at the start of each sequence. Note however, that the optimization algorithm can set the flip angle to 0 degrees, thus effectively creating a delay.

As a measure of optimality, the Cramér-Rao Lower Bound (CRLB) \cite{Kay1993} was used. This measure expresses the minimum variance of a set of estimated parameters, in this case $T_1$ and $B_1^+$, obtained using an unbiased estimator, here the MRF reconstruction. The CRLB has previously been used to find optimal parameters for an MRI experiment, as well as for MRF in particular \cite{Teixeira2018, Zhao2019, Lee2019, Asslander2019}.

The measurements are assumed to be subject to white Gaussian noise:
\begin{equation} \label{eq:GaussNoise}
    s_n(\theta,\alpha)=M_n(\theta,\alpha)+w_n
\end{equation}
where $s_n$ is the measured signal intensity in the $n$\textsuperscript{th} set, $M_n$ is the normalized and averaged signal intensity of the $n$\textsuperscript{th} set as calculated by the same simulator used for the dictionary, $\theta\in\mathbb{R}^p$ is the vector of all estimated parameters ($T_1$ and $B_1^+$ in this case), $\alpha\in\mathbb{R}^q$ is the vector containing all flip angles in the sequence, and $w_n\sim\mathcal{N}(0,\sigma^2)$ is normally distributed noise with standard deviation $\sigma$. Since the fingerprints are normalized, the standard deviation of the measurement noise should be scaled identically to obtain the right $\sigma$. The Fisher information matrix (FIM), which can be used to obtain the CRLB \cite{Kay1993}, can be written as:
\begin{equation} \label{eq:FIM}
    I(\theta,\alpha)=\frac{1}{\sigma^2}\sum_{n=1}^{N}{J_n(\theta,\alpha)^T \cdot J_n(\theta,\alpha)}
\end{equation}
with $I(\theta,\alpha)\in\mathbb{R}^{p\times p}$ the Fisher information matrix, $\sigma$ the standard deviation scaled according to the normalization of the fingerprint, and $J_n(\theta,\alpha)=\frac{\partial M_n(\theta,\alpha)}{\partial \theta}\in\mathbb{R}^{1\times p}$ the Jacobian of the signal $M_n$ with respect to the estimated parameters $\theta$. Taking the inverse of the FIM gives a matrix, whose diagonal elements indicate the CRLBs of the estimated parameters given the flip angle sequence \cite{Kay1993}:
\begin{equation} \label{eq:CRLB}
    Var(\hat{\theta_i} | \alpha) \geq [I^{-1}(\theta,\alpha)]_{i,i}
\end{equation}
These variances were normalized by the square of the true $T_1$ and $B_1^+$ values respectively, to give the coefficient of variation (COV) for both parameters. To make the MRF sequence sensitive to both $T_1$ and $B_1^+$, the sum of these two relative CRLBs (the trace of the covariance matrix) was minimized.

The CRLB of a parameter gives an estimate of the sensitivity around a specific parameter value. Since we want the MRF experiment to be sensitive to a range of parameter values, the relative CRLBs for $N_l$ different combinations of $T_1$ and $B_1^+$ values were calculated and averaged. Furthermore, all flip angles were limited to 60 degrees, corresponding to the peak transmit voltage for an average subject when using an apodized sinc-shaped RF-pulse of 1.5 ms with a time-bandwidth factor of 8. The final optimization problem is given by Eq. \ref{eq:Optimization}.
\begin{equation} \label{eq:Optimization}
    \begin{array}{rll}
        \displaystyle \min_{\alpha} & \multicolumn{2}{l}{\displaystyle \frac{1}{N_l} \sum_{l=1}^{L} \textnormal{tr}(W \cdot I^{-1}(\theta_l,\alpha))} \\
        \textnormal{s.t.} & 0 \degree \leq \alpha_n \leq 60 \degree, & 1 \leq n \leq q
    \end{array}
\end{equation}
Here $\theta_l\in\mathbb{R}^p$ is the $l$\textsuperscript{th} parameter combination, $W\in\mathbb{R}^{p\times p}$ is a diagonal matrix with weighting factors for the parameters, $\textnormal{tr}(\cdot)$ denotes the trace of a matrix, $\alpha_n$ is the $n$\textsuperscript{th} flip angle in the sequence, and $q$ is the total number of flip angles. We used $W=\textnormal{diag}([1/(T_1)^2, 1/(B_1^+)^2])$, which normalized the variances of both parameters. 

During optimization, the number of flip angles was fixed. To investigate the influence of different sequence lengths, the sequence was optimized for 300, 600, 900, and 1800 flip angles, using $T_1$ values between 500 ms and 1500 ms in steps of 250 ms, and relative $B_1^+$ values of 0.75, 1.0, and 1.25, giving $N_l = 15$. The sequences were initialized with smooth random flip angle patterns with values between 0 and 10 degrees. The optimization was repeated for 20 different initializations. There was no explicit delay time between subsequent repetitions of the sequence. However, note that the optimization algorithm was able to set the flip angles to negligible low values, thus effectively creating a delay that allowed the magnetization to relax towards equilibrium.

The optimization problem given in Eq. \ref{eq:Optimization} was implemented in MATLAB (The MathWorks Inc., Natick, MA, USA). Automatic differentiation \cite{Baydin2018, Lee2019, Asslander2019}, implemented using the CasADI toolbox \cite{Andersson2019}, was used to calculate the Jacobian in Eq. \ref{eq:FIM}, as well as the gradient and Hessian-times-vector product of the objective function in Eq. \ref{eq:Optimization}. The minimization problem was solved with the MATLAB function fmincon. The code used to optimize the flip angles is available on our bitbucket: \url{https://bitbucket.org/MaxvRiel/free-breathing-mr-fingerprinting/}.

\begin{figure}[!t]
    \centering
    \includegraphics[width=7.5cm]{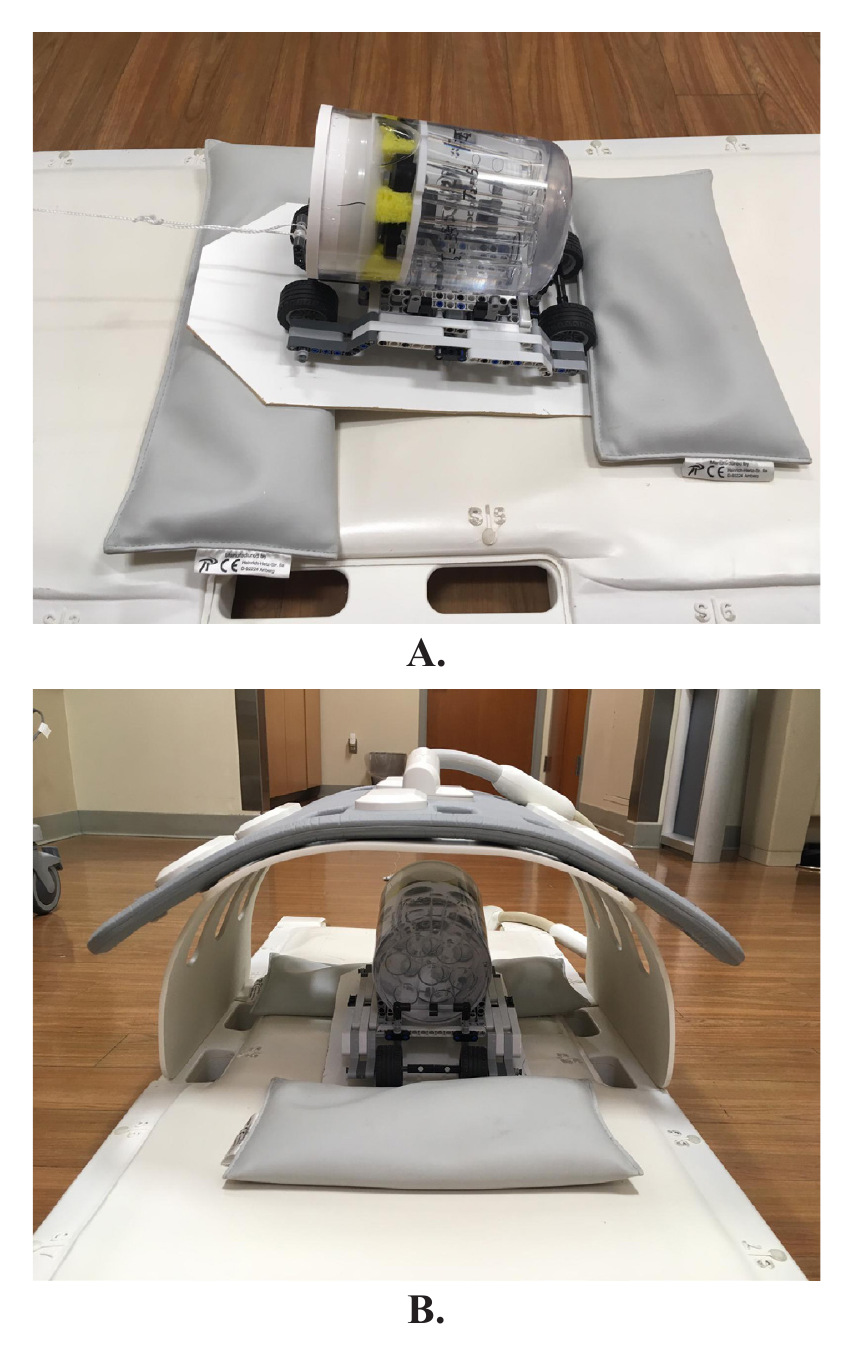}
    \caption{Experimental setup. \textbf{A}: The phantom placed on a movable cart. This cart is placed on a ramp. Under the influence of gravity, the cart moves to the right, while a rope connected to a motor can pull the cart to the left. \textbf{B}: The 18-channel body coil that was used to acquire the data was placed over the movable phantom.}
    \label{fig:ExperimentSetup}
\end{figure}

\subsection{Imaging Experiments}

\begin{table*}[!t]
	\renewcommand{\arraystretch}{1.2}
	\caption{Image acquisition parameters used in the phantom and in vivo experiments.}
	\label{table:ImagingParameters}
	\centering
	\begin{tabular}{c|c|c|c}
		\hline
		\textbf{Parameter} & \textbf{Value} & \textbf{Value} & \textbf{Unit} \\
		& \textbf{(phantom)} & \textbf{(in vivo)} & \\
		\hline\hline
		Repetition time (TR) & \multicolumn{2}{|c|}{5.0} & ms \\
		\hline
		Echo time (TE) & \multicolumn{2}{|c|}{2.4} & ms \\
		\hline
		Inversion time (TI) & \multicolumn{2}{|c|}{10} & ms \\
		\hline
		Number of partitions & \multicolumn{2}{|c|}{30} & - \\
		\hline
		Acquisition time & \multicolumn{2}{|c|}{4:30} & min:sec \\
		\hline
		Resolution & 1.0$\times$1.0$\times$6.0 & 1.3$\times$1.3$\times$3.0 & mm\textsuperscript{3} \\
		\hline
		Field of view & 256$\times$256$\times$180 & 420$\times$420$\times$90 & mm\textsuperscript{3} \\
		\hline
	\end{tabular}
\end{table*}

All experiments were performed on a clinical 3T MRI scanner (Prisma, Siemens Healthineers, Erlangen, Germany). A phantom containing 7 glass tubes with different $T_1$ values was placed on a cart made from LEGO\textsuperscript{\textregistered} bricks (The Lego Group, Billund, Denmark). This cart was placed on a slope of approximately 7 degrees and connected to a motor just outside the scanner room, which was controlled using the RWTH Mindstorms NXT Toolbox for MATLAB (RWTH Aachen University, Aachen, Germany). An 18-channel body coil (Siemens) was placed over the phantom and the cart (Figure \ref{fig:ExperimentSetup}). The phantom was scanned while stationary and while moving back and forth with frequencies of 0.22, 0.24, and 0.30 Hz, which are within the range of normal breathing frequencies in adults \cite{Lindh2010DelmarsCompetencies}. All scans were performed with both ordering schemes. A previously validated 2D MRF implementation \cite{Cloos2019} was used to obtain reference $T_1$ values in the absence of motion.

In order to keep the same acquisition time of 4 minutes and 30 seconds for the different flip angle trains, the number of shots was increased for shorter sequences. This resulted in 6, 3, 2, and 1 shots for the sequences with 300, 600, 900, and 1800 TRs, respectively.

An abdominal scan was performed on one healthy volunteer after obtaining written informed consent. The study was approved by our institutional review board. The same scanner and body coil were used as during the phantom scan. The volunteer was instructed to breathe normally, but to avoid taking deep breaths (by sighing or yawning). The parameters used for both experiments are summarized in Table \ref{table:ImagingParameters}. 

\subsection{Image Reconstruction}

After acquiring all the data, the receive channels were compressed to 5 (phantom) or 10 (in vivo) virtual coils for every slice independently, retaining more than 98.5\% of the energy in all datasets. A Fast Fourier Transform (FFT) was performed in the phase encoding direction. A rank-5 approximation of the k-space data was made using the same 5 singular vectors computed for the dictionary compression \cite{McGivney2014SVDDomain}. Next, every partition was reconstructed separately using the Nonuniform Fast Fourier Transform (NUFFT) \cite{Fessler2003}, and the virtual coil channels were combined using a matched-filter reconstruction \cite{Roemer1990}. 

For every voxel, the dot product between the fingerprint and the dictionary was maximized to identify the $T_1$ and $B_1^+$ values associated with that voxel. The Euclidean norm corresponding to the (unnormalized) dictionary entry was used to estimate the PD. The image reconstruction, dictionary construction and dictionary matching were all implemented in MATLAB (The MathWorks Inc., Natick, MA, USA). The code used to reconstruct the parameter maps is available at \url{https://bitbucket.org/MaxvRiel/free-breathing-mr-fingerprinting/}.

\subsection{Image Analysis}

\begin{figure}[!t]
    \centering
    \includegraphics[width=7cm]{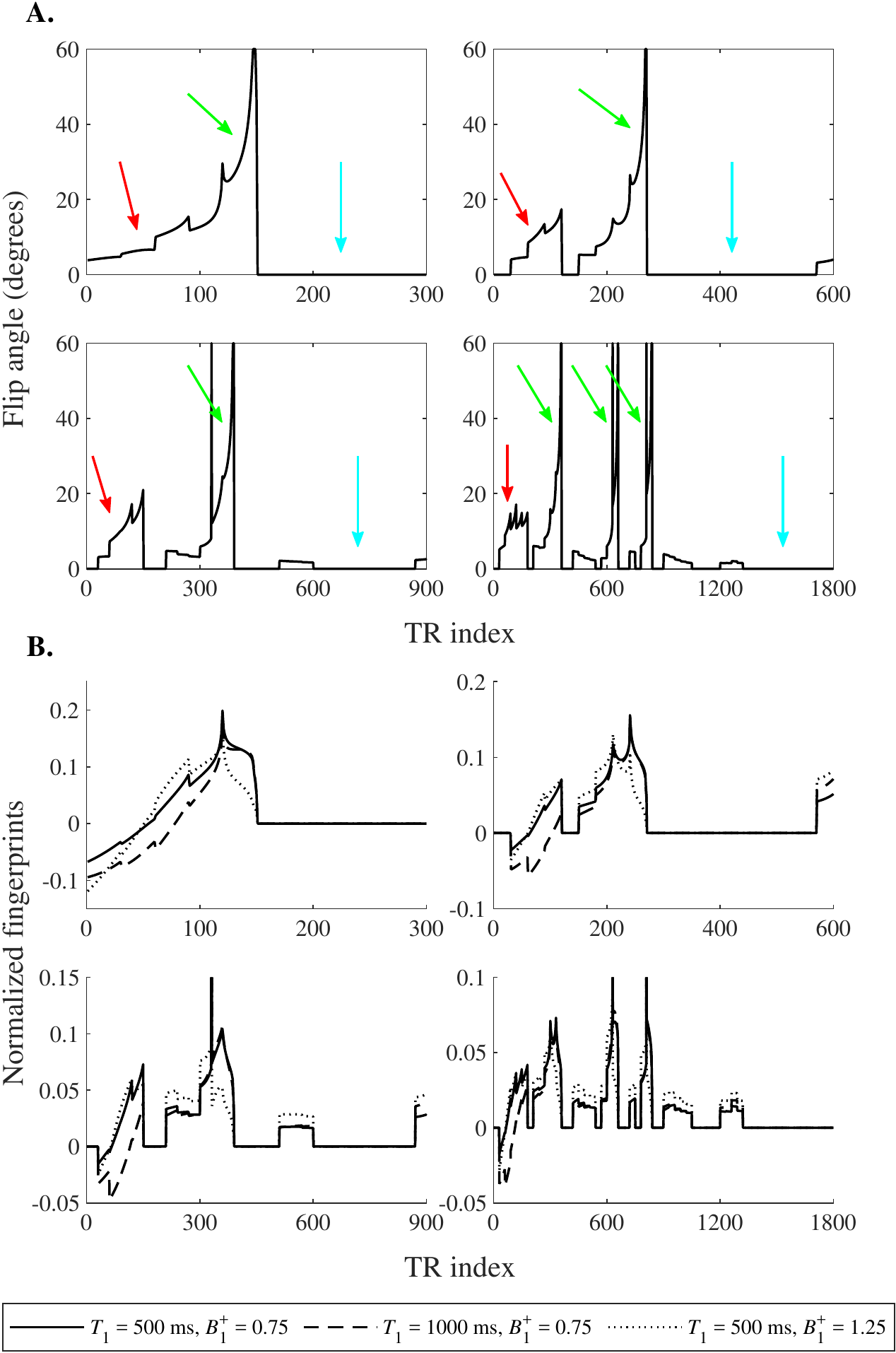}
    \caption{Results of the flip angle optimization. \textbf{A}: Optimized flip angle patterns for 300, 600, 900, and 1800 TRs (left to right, top to bottom). Note the smoothly varying flip angles within each set. Indicated are the main $T_1$-encoding part (red arrows), the main $B_1^+$-encoding part (green arrows), and the delay (light blue arrows). \textbf{B}: Normalized fingerprints for different combinations of $T_1$ and $B_1^+$ values for the same four flip angle patterns.}
    \label{fig:OptimizedFAPatterns}
\end{figure}

\begin{figure*}[!t]
    \centering
    \includegraphics[width=12cm]{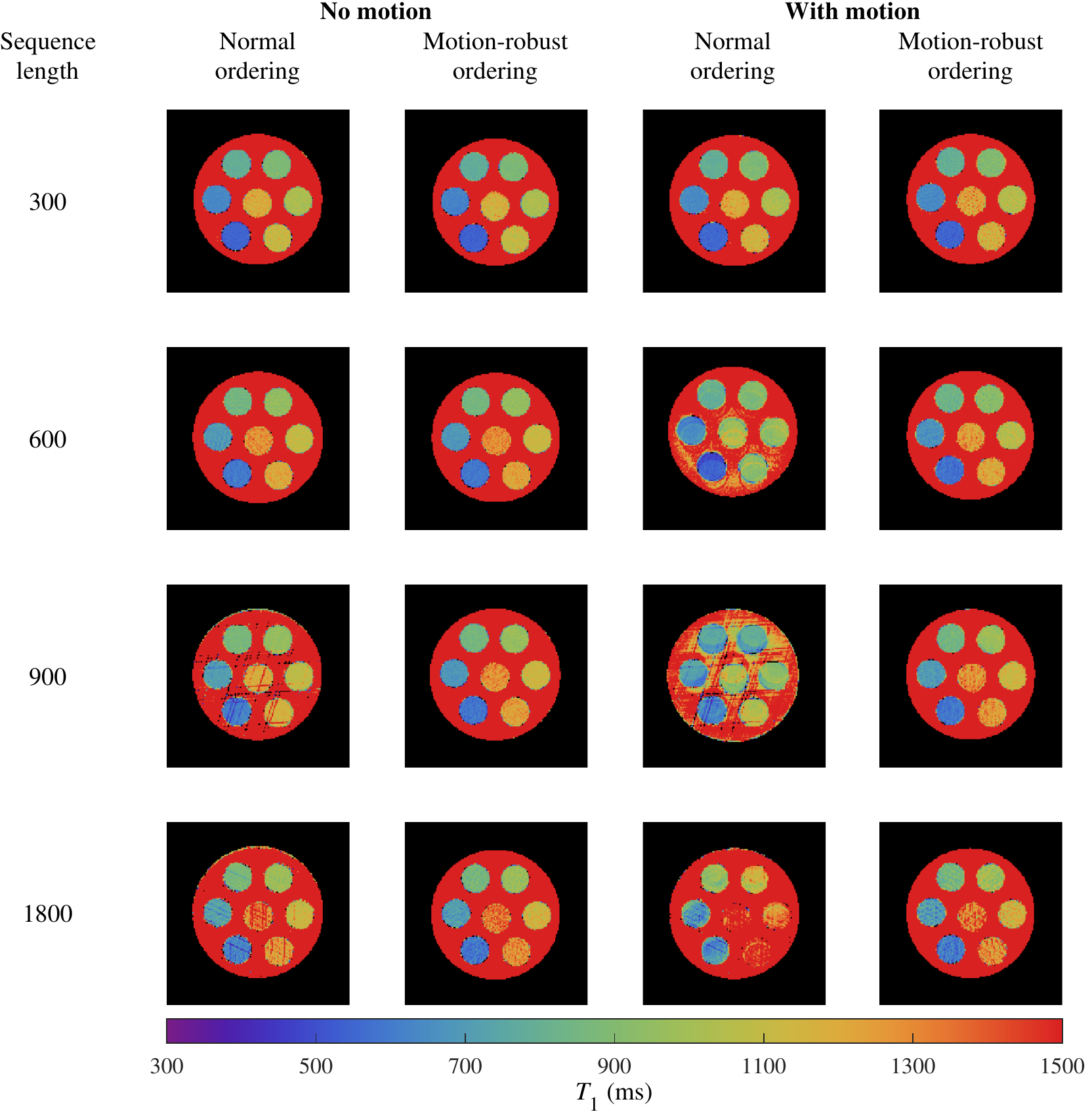}
    \caption{$T_1$ maps of the phantom scan without (left) and with (right) motion, for both ordering schemes, and for all four sequence lengths for which the flip angles were optimized. For the maps with motion, the motion speeds with the most severe artefacts were selected for each sequence separately. Note the severe motion-related artefacts visible when using the \textit{normal ordering}, which are greatly reduced by using the \textit{motion-robust ordering}.}
    \label{fig:PhantomResult}
\end{figure*}

\begin{figure*}[!t]
    \centering
    \includegraphics[width=12cm]{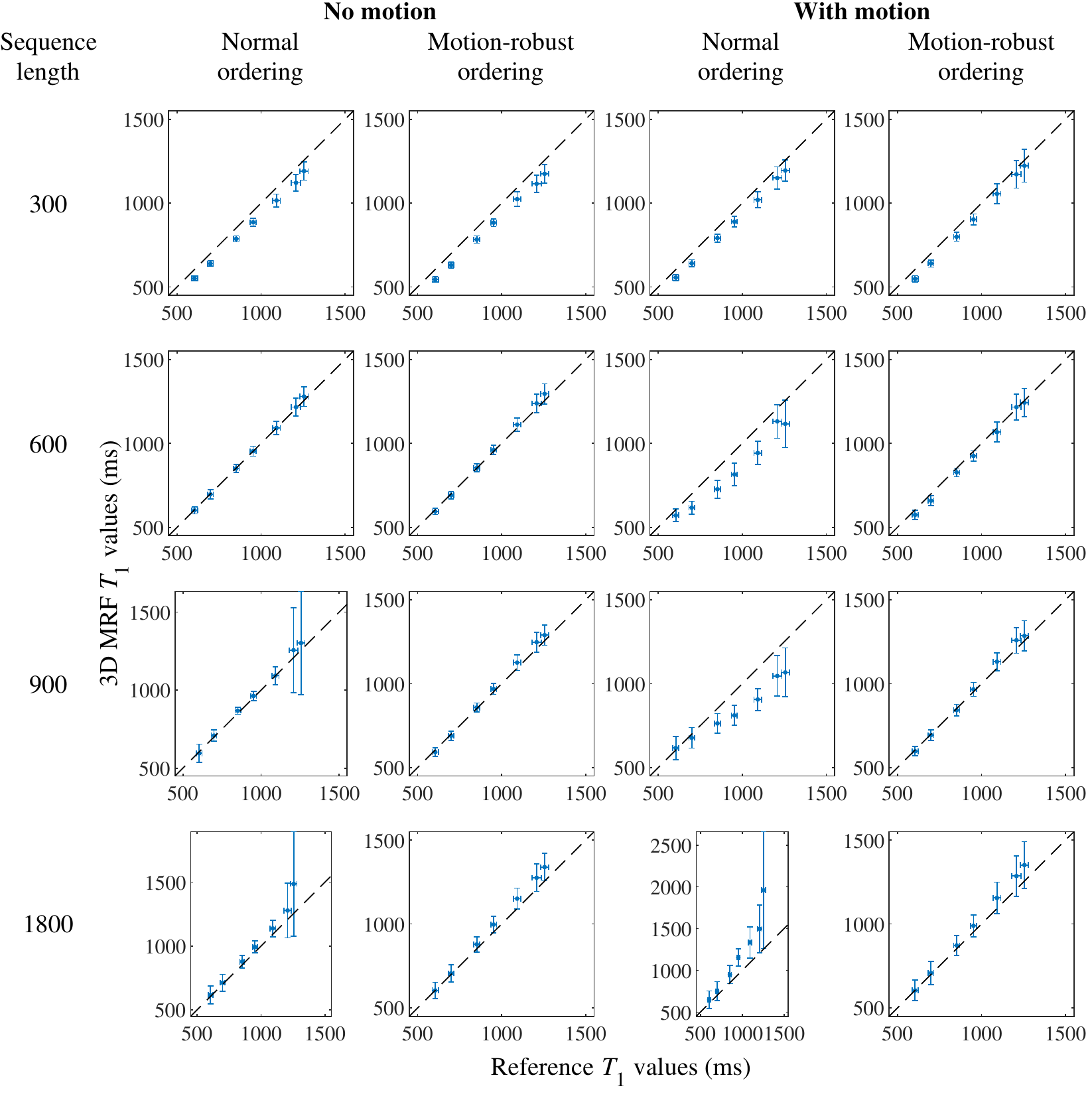}
    \caption{Correlation plots between the estimated $T_1$ values (vertical axis) and the corresponding $T_1$ values from the reference scan (horizontal axis) for each map in Figure \ref{fig:PhantomResult}. Each point indicates the mean $T_1$ values of both scans within one single tube, with the standard deviation depicted as error bars. The dashed line is the identity line. Note the increased deviation from the identity line for the \textit{normal ordering} compared to the \textit{motion-robust ordering}.}
    \label{fig:PhantomCorrelations}
\end{figure*}

In order to analyze the parameter maps of the moving phantom, a region of interest (ROI) was drawn in each of the 7 tubes. The mean and standard deviation of all estimated parameters were calculated for each tube. The estimated $T_1$ values were validated against those from the reference scan.

\section{Results}
\subsection{Flip Angle Patterns}

The flip angles optimized for different train lengths (Figure \ref{fig:OptimizedFAPatterns}) showed similar patterns. Within each set of 30 consecutive TRs (the number of partitions), the flip angles generally increase smoothly, which results in smoothly varying signal intensities within each set. The flip angles only show large jumps between sets. 

Furthermore, the flip angles of the first few sets in all sequences follow a similar pattern, which occurs around the zero-crossing of the inverted magnetization. Hence, it is mostly dependent on $T_1$ and mainly serves to encode $T_1$ in the fingerprint (solid and dashed lines in Figure \ref{fig:OptimizedFAPatterns}B). 

Next, there is a recurring pattern which consists of a single high flip angle, followed by a set of increasing flip angles. This pattern is particularly visible in the sequences with 900 and 1800 TRs but is also present in the sequences with 300 and 600 TRs. It mainly serves to provide $B_1^+$-sensitive signals (solid and dotted lines in Figure \ref{fig:OptimizedFAPatterns}B). The single high flip angle partly saturates the signal, with the amount of saturation depending on the $B_1^+$ value. The succeeding excitations then provide the $B_1^+$-dependent signal for the fingerprint. Note that this pattern is repeated several times in the sequence with 1800 TRs.

Since the magnetization is saturated after this $B_1^+$-encoding pattern, any subsequent excitations would not generate much signal. The optimization resulted in multiple segments with flip angles less than 1.5 degree, which were zeroed as these would be dominated by noise. During these segments, the magnetization can recover before the next repetition of the sequence.

\subsection{Phantom Scan}

\begin{figure*}[!t]
    \centering
    \includegraphics[width=12cm]{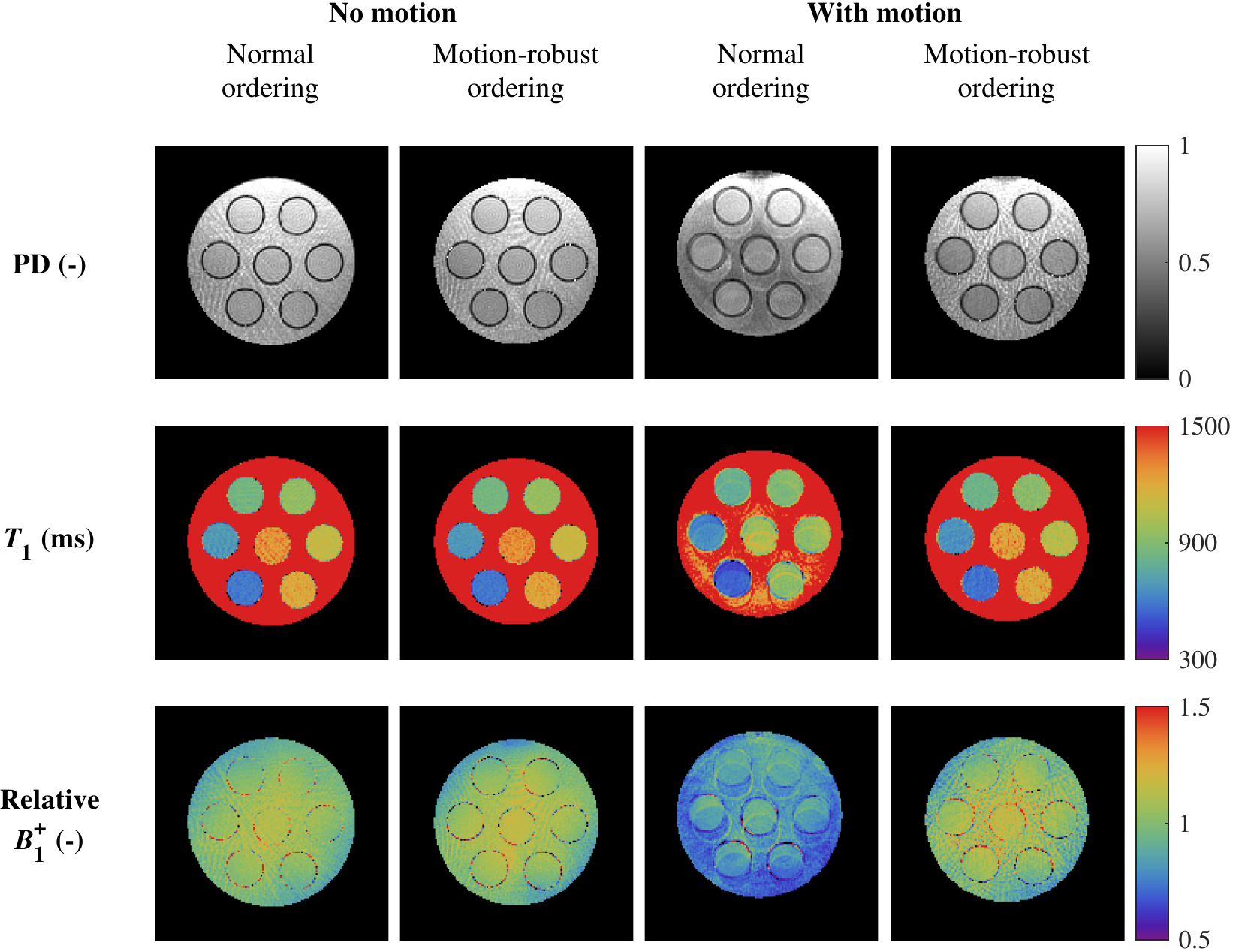}
    \caption{Quantitative PD (top row), $T_1$ (middle row), and $B_1^+$ (bottom row) maps of the phantom, both without (left) and with (right) motion, and with both ordering schemes, using the optimized sequence with 600 flip angles and 3 shots. Note the motion-related artefacts in the parameter maps acquired with the \textit{normal ordering}.}
    \label{fig:PhantomAllParams}
\end{figure*}

\begin{figure*}[!t]
    \centering
    \includegraphics[width=12cm]{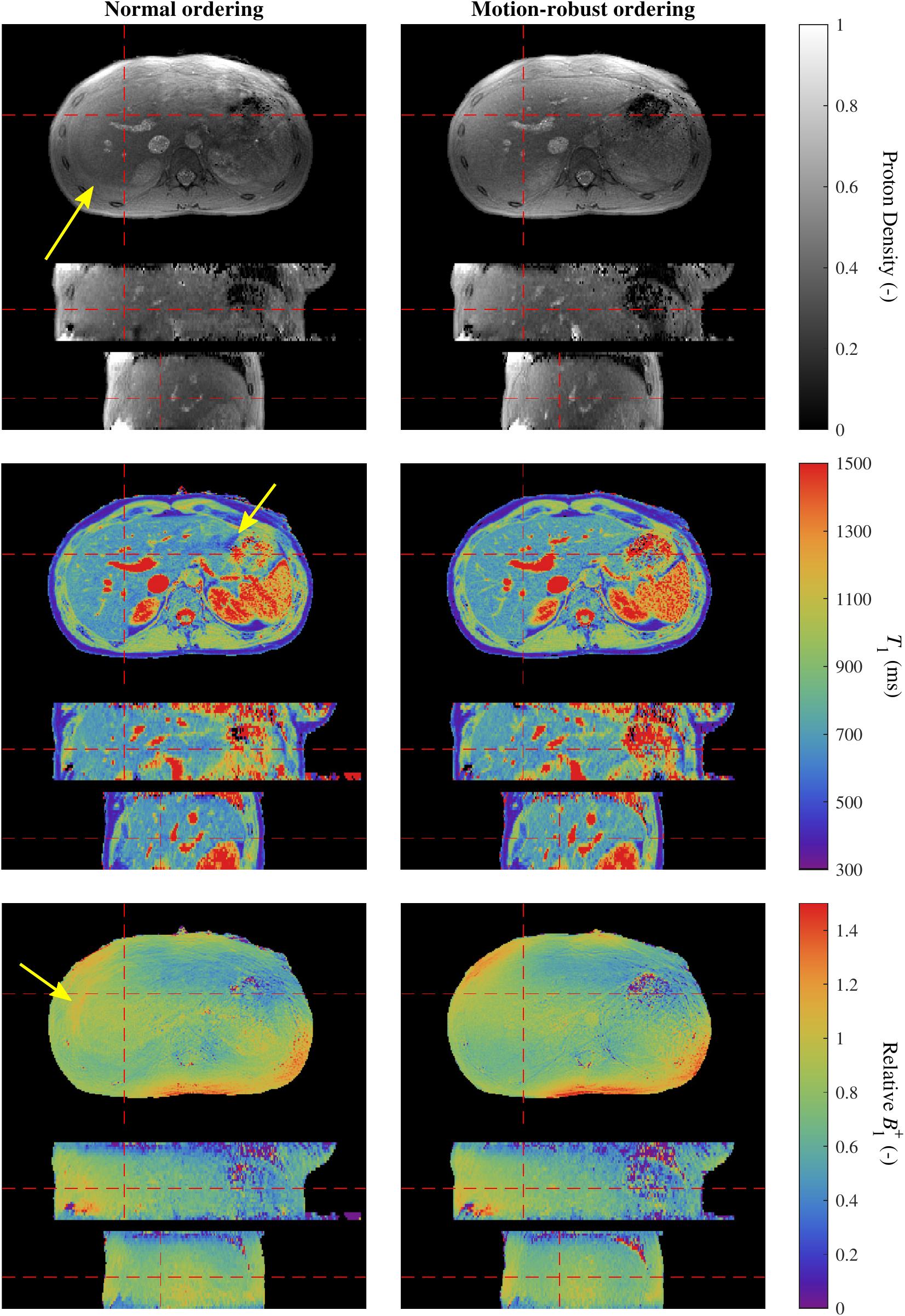}
    \caption{In vivo quantitative PD (top row), $T_1$ (middle row), and $B_1^+$ (bottom row) maps for the \textit{normal ordering} (left column) and the \textit{motion-robust ordering} (right column), using the optimized sequence with 600 flip angles and 3 shots. The \textit{motion-robust ordering} reveals more details in the $T_1$ map and removes the motion-related artefacts visible in the maps of all three parameters (indicated with yellow arrows).}
    \label{fig:InVivoResult}
\end{figure*}

The $T_1$ maps from the phantom scan for all four optimized sequences can be seen in Figure \ref{fig:PhantomResult}. For PD, $T_1$, and $B_1^+$ maps, acquired using all tested motion speeds, see Supplementary Figures S1-S8. The sequence with 300 flip angles and 6 shots slightly underestimated higher $T_1$ values. The longest flip angle train (1800 flip angles, 1 shot) showed large variability in the $T_1$ estimates as well. This is also visible when looking at the correlation between the estimated and the reference $T_1$ values (Figure \ref{fig:PhantomCorrelations} and Supplementary Figure S9). The other two sequences showed a better compromise between the number of shots and encoding capability. 

The sequence with 600 flip angles was selected to be investigated further, as this sequence showed the highest accuracy for the $T_1$ estimation. In Figure \ref{fig:PhantomAllParams}, the estimated parameter maps for this sequence with both ordering schemes and both with and without motion are shown. Without motion, both ordering schemes showed good agreement with the reference scan. In the presence of motion however, the \textit{normal ordering} showed notable motion artefacts in all parameter maps, and a large deviation from the identity line in the $T_1$ correlation plots. The normalized root mean square error, comparing the estimated $T_1$ values of the tubes measured in the absence of motion with the measurements including motion, was also reduced from 0.17 for the \textit{normal ordering} to 0.05 for the \textit{motion-robust ordering}.

The severity of the observed motion artefacts depends on the frequency of the motion. Presented here are the results obtained with the speeds that caused the most severe artefacts when using the \textit{normal ordering}. Nevertheless, it should be noted that the \textit{motion-robust ordering} performed well independent of the speed of the motion (Supplementary Figure S4).

\subsection{In Vivo Scan}

Since the standard deviation of the $T_1$ values obtained using the sequence with 600 flip angles was the lowest, this sequence was evaluated in vivo (Figure \ref{fig:InVivoResult}). The estimated $T_1$ values inside the liver were around 700 ms, comparable to values found in literature \cite{Chan2012, Obmann2019LiverStiffness}. Note the variations of the excitation field strength in the $B_1^+$ maps, which underline the importance of taking $B_1^+$ inhomogeneities into account. 

The \textit{motion-robust ordering} resulted in a sharper $T_1$ map, where the vessels in the liver are much more visible, when compared to the \textit{normal ordering}. Furthermore, there are some severe motion artefacts visible in the maps acquired using the \textit{normal ordering} (yellow arrows in Figure \ref{fig:InVivoResult}). These artefacts are no longer visible when using the \textit{motion-robust ordering}. 

The raw in vivo data is available at \url{https://data.mendeley.com/datasets/txpwybnt5p/}.

\section{Discussion}
In this work, we have demonstrated a 3D MRF sequence for free-breathing abdominal imaging. In addition, excitation field inhomogeneities were accounted for by including $B_1^+$ in the parameter estimation. Both phantom and in vivo scans showed that motion artefacts are reduced when using the \textit{motion-robust ordering}. Because the MR signal in an MRF experiment is not in steady-state, the signal intensity within each set is not constant in the phase encoding direction when using the \textit{motion-robust ordering} scheme. An additional constraint could be added to the optimization problem to enforce signal smoothness. However, the absence of such a constraint did not result in significant artefacts in the parameter maps when using the \textit{motion-robust ordering}.

All optimized flip angle trains included a segment without any data acquisition. Other MRF implementations used a delay time between flip angle segments \cite{Ma2013, Cloos2016} to allow the magnetization to recover. In this work, the delay time was not set explicitly. Instead, the total time of data acquisition was set by determining the number of TRs. The optimization algorithm could thus find the optimal compromise between delay time and data acquisition. In particular, when using more TRs, the optimization resulted in longer delays.

One important thing to note is that the phantom was moved at different speeds, but not all speeds showed the motion artefacts when using the \textit{normal ordering} (Supplementary Figures S1-S8). This was probably caused by interference between the motion pattern and the timing of the sequence, which repeats the flip angle train every few seconds (the exact duration depends on the number of flip angles in the sequence). In other words, if the phantom happens to be in the same place at the start of each flip angle train, the motion artefacts are minimal. However, human subjects have a wide range of breathing frequencies. Moreover, these breathing rhythms can be irregular. Consequently, as demonstrated in the in vivo scan, the proposed motion robust ordering is important to prevent motion artefacts.

Different sequence lengths were compared by using the moving phantom. Shorter sequences are less flexible in their encoding but are faster to acquire per repetition. To compare different sequence lengths, we used more shots (and thus more repetitions) for shorter sequences to keep the total acquisition time constant. For the shortest sequence, the parameter estimations were less accurate, especially for the tubes with longer $T_1$ values. Most likely, the duration of one repetition was too short to observe the slow dynamics of the long $T_1$ samples. By increasing the sequence length to 600 and 900 TRs, the estimated quantitative values were improved. For the longest sequence with 1800 TRs however, the accuracy decreased again, presumably because the undersampling artefacts were more pronounced in the reconstructed images due to the lower number of shots. We found an optimal compromise between encoding capability and acquisition speed around 600 TRs, corresponding to a 3000 ms interval between two inversion pulses. Note that our optimization routine sets some of these flip angles to zero to create delays.

Although considerable effort went into optimizing the sequence, there is still room for further improvements. It is possible that the insertion of inversion or saturation pulses in the sequence could improve the encoding capability. Currently, the optimization algorithm is unable to do this, since the peak flip angle is limited to 60 degrees to account for the peak power the system can provide. Additionally, it could be investigated whether regularization could improve the optimization algorithm. For example, the signal differences within one set could be minimized by adding a term in Eq. \ref{eq:Optimization} to reduce blurring in the images \cite{Deichmann2000}. Furthermore, in this work only spoiled gradient echoes were used. The inclusion of Fast Imaging with Steady State Precession (FISP) segments, as used before in MRF \cite{Cloos2016, Jiang2015}, could add valuable clinical information. However, this would increase the complexity for both the dictionary construction and the optimization algorithm. Moreover, FISP segments are notoriously sensitive to motion \cite{Yu2018}.

Besides the sequence itself, the reconstruction process could be improved as well. Low-rank methods \cite{Asslander2018LowFingerprinting} or parallel imaging techniques \cite{Liao20173DReconstruction} could be used to reduce the undersampling artefacts and improve the quantitative maps. These methods could also be used to increase the number of partitions, thereby increasing the field of view. Additionally, previous work has shown that it is possible to correct for rigid motion in brain MRF \cite{Cruz2017, Mehta2018, Xu2019, Cruz2019}. A similar approach could possibly reduce the motion-related artefacts even further, resulting in better images and consequently more accurate parameter maps. However, motion in the abdomen is nonrigid. Therefore, further studies will be needed to explore if such a motion correction step is feasible.

The goal of this work was to introduce a new free-breathing MRF sequence. To prove the robustness of the proposed sequence, more subjects will have to be scanned. Finally, to investigate the performance of the sequence with different pathologies, several clinical scans with patients will have to be performed as well.

\section{Conclusion}
A free-breathing MR Fingerprinting sequence was demonstrated for $B_1^+$-robust quantitative abdominal imaging. Four different flip angle patterns were optimized. A movable phantom was used to validate the sequences. The flip angle train with 600 TRs provided the best trade-off between $T_1$ encoding power and sampling density. In vivo measurements confirmed the advantage of the motion-robust k-space ordering. With this free-breathing MRF implementation it is possible to collect crisp PD images and clean $T_1$ maps of the abdomen at a clinically usable resolution within 5 minutes.

\section*{Acknowledgements}
The research reported in this publication was supported by the NIH/NIBIB grant R01 EB026456, NIH/NIAMS grant R01 AR070297, and performed under the rubric of the Center for Advanced Imaging Innovation and Research, an NIBIB Biomedical Technology Resource Center (P41 EB017183). Furthermore, we would like to express our gratitude towards the Holland Scholarship from the Dutch Ministry of Education, Culture and Science, as well as the Professor Huson award and the BMT Bachelorbeurs from the Eindhoven University of Technology for supporting this project.

\balance

\bibliographystyle{unsrt}
\bibliography{references}

\end{document}


\clearpage

\maketitle

\begin{figure}[!t]
    \centering
    \includegraphics[width=12.7cm]{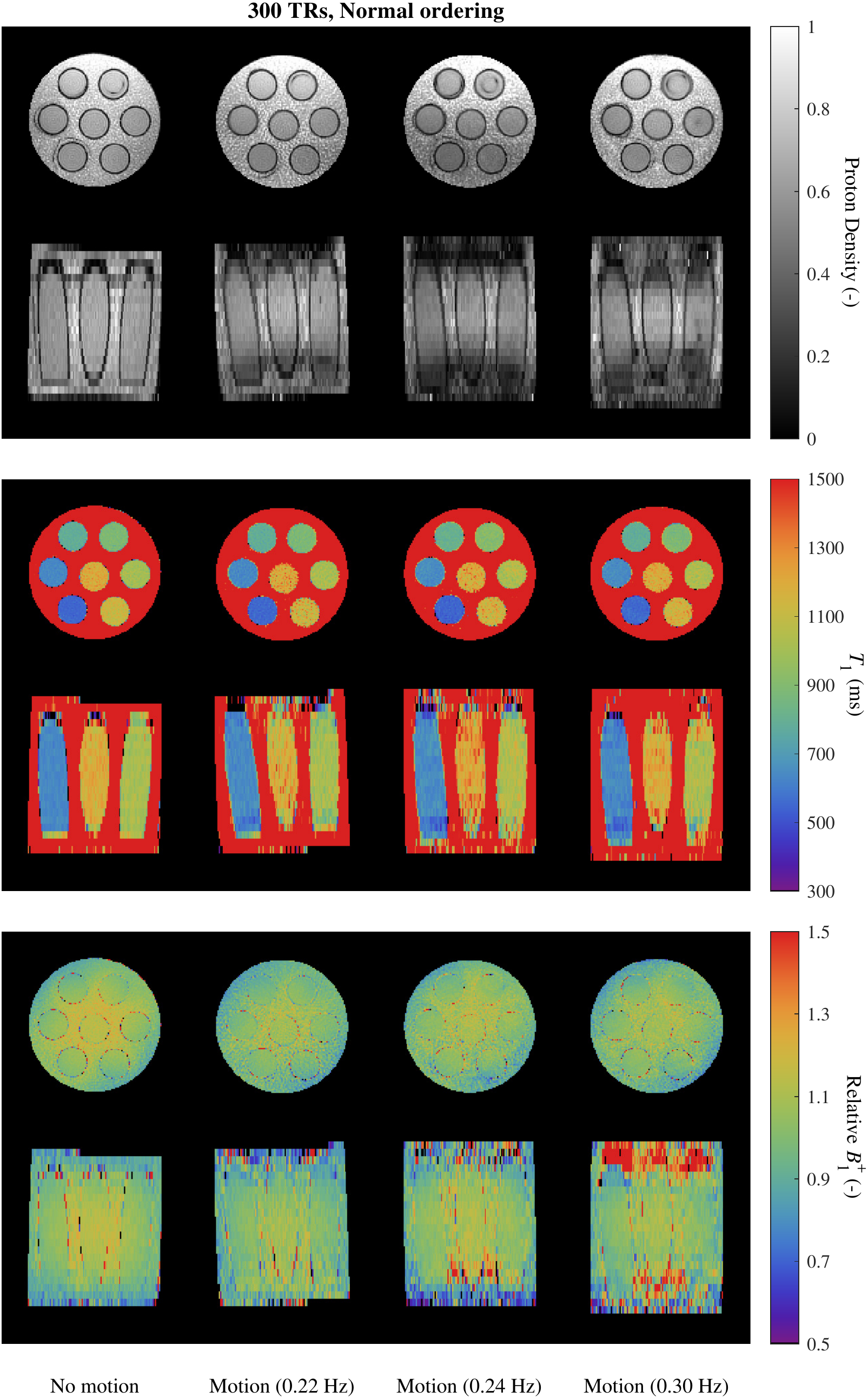}
    \caption{Quantitative Proton Density (top row), $T_1$ (middle row), and $B_1^+$ (bottom row) maps for the sequence with 300 TRs and the \textit{normal ordering}. For each map, both a transverse (top) and a coronal (bottom) slice is shown.}
    \label{SIfig:300_norm}
\end{figure}

\begin{figure}[!t]
    \centering
    \includegraphics[width=12.7cm]{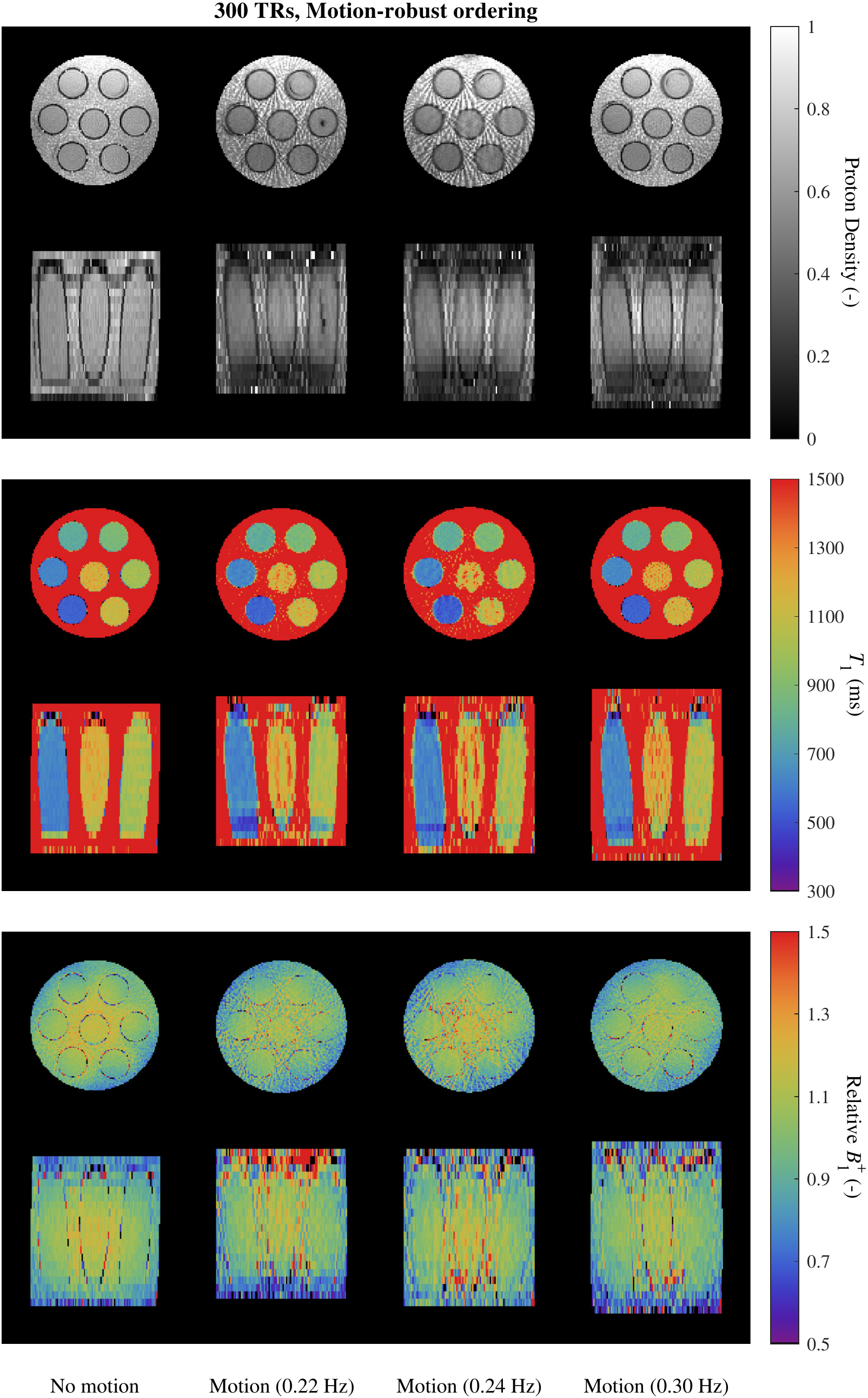}
    \caption{Quantitative Proton Density (top row), $T_1$ (middle row), and $B_1^+$ (bottom row) maps for the sequence with 300 TRs and the \textit{motion-robust ordering}. For each map, both a transverse (top) and a coronal (bottom) slice is shown.}
    \label{SIfig:300_mr}
\end{figure}

\begin{figure}[!t]
    \centering
    \includegraphics[width=12.7cm]{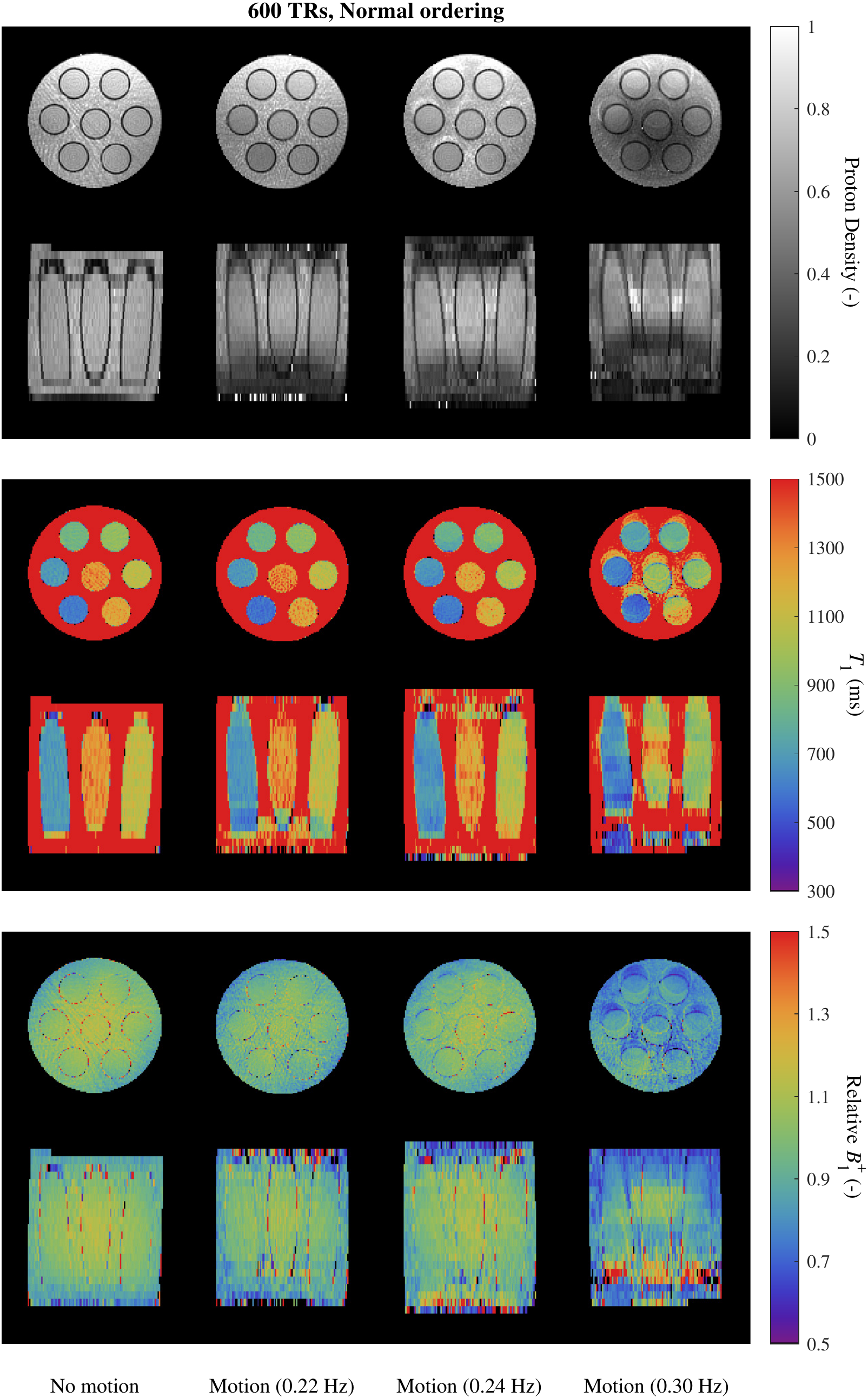}
    \caption{Quantitative Proton Density (top row), $T_1$ (middle row), and $B_1^+$ (bottom row) maps for the sequence with 600 TRs and the \textit{normal ordering}. For each map, both a transverse (top) and a coronal (bottom) slice is shown.}
    \label{SIfig:600_norm}
\end{figure}

\begin{figure}[!t]
    \centering
    \includegraphics[width=12.7cm]{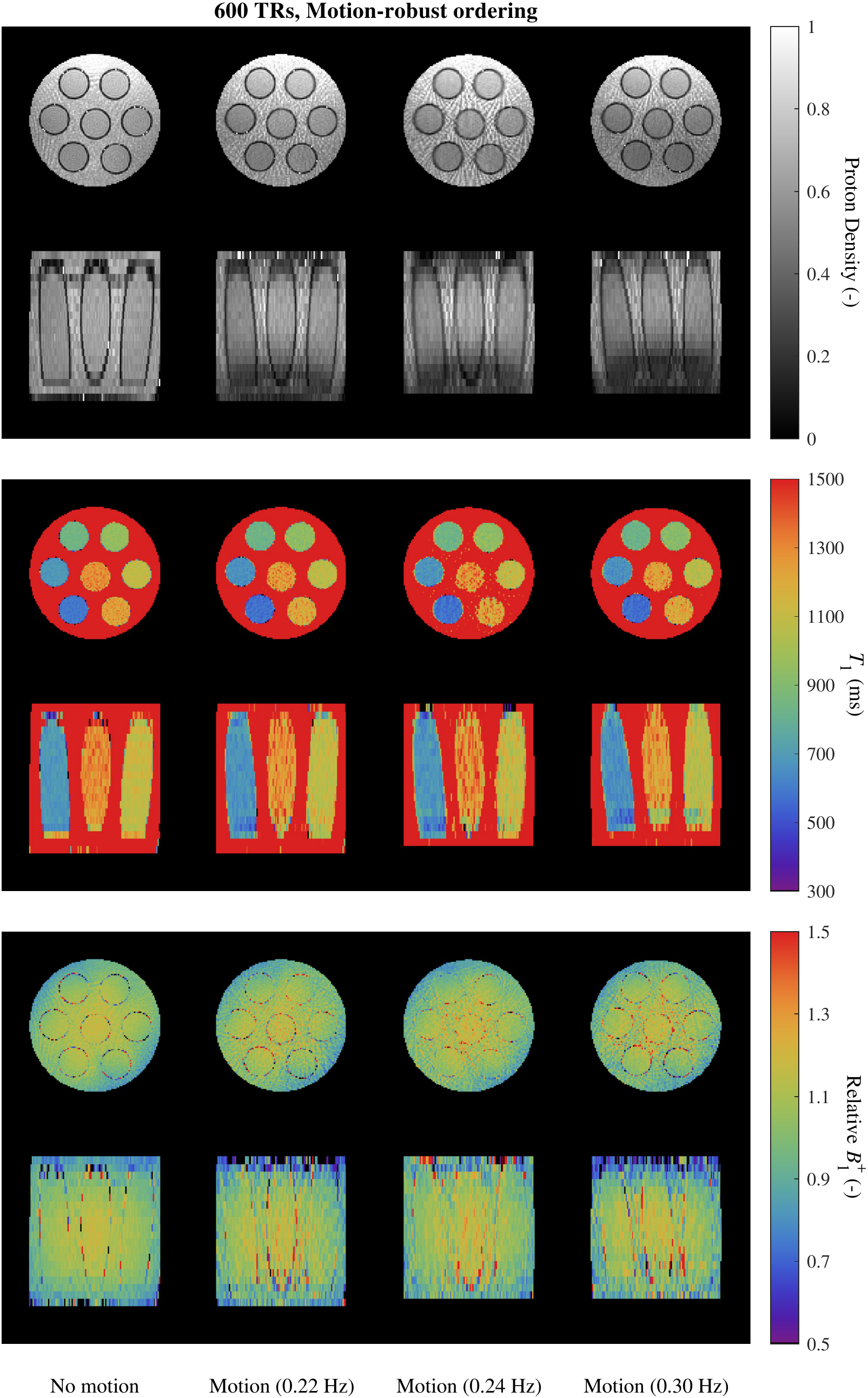}
    \caption{Quantitative Proton Density (top row), $T_1$ (middle row), and $B_1^+$ (bottom row) maps for the sequence with 600 TRs and the \textit{motion-robust ordering}. For each map, both a transverse (top) and a coronal (bottom) slice is shown.}
    \label{SIfig:600_mr}
\end{figure}

\begin{figure}[!t]
    \centering
    \includegraphics[width=12.7cm]{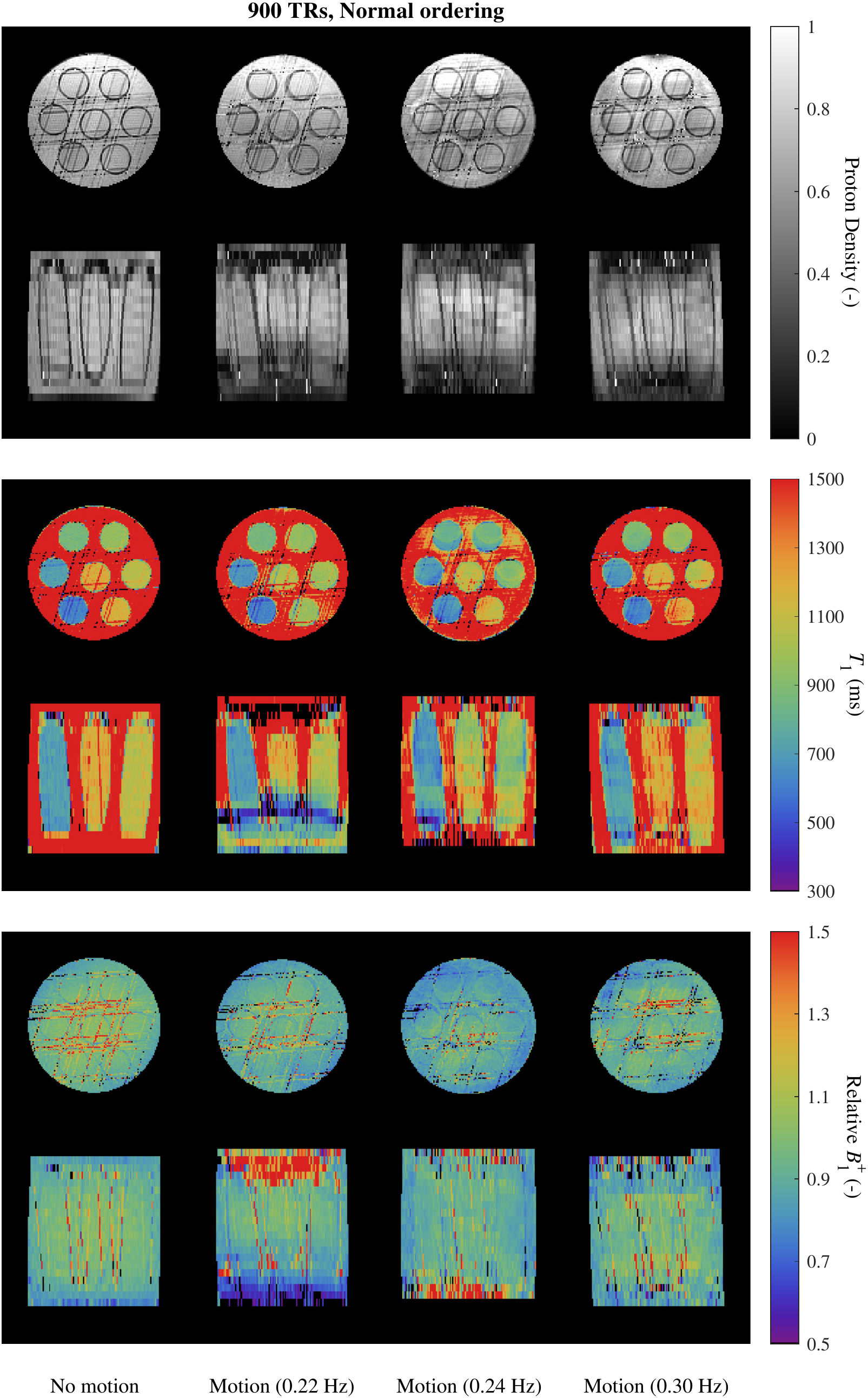}
    \caption{Quantitative Proton Density (top row), $T_1$ (middle row), and $B_1^+$ (bottom row) maps for the sequence with 900 TRs and the \textit{normal ordering}. For each map, both a transverse (top) and a coronal (bottom) slice is shown.}
    \label{SIfig:900_norm}
\end{figure}

\begin{figure}[!t]
    \centering
    \includegraphics[width=12.7cm]{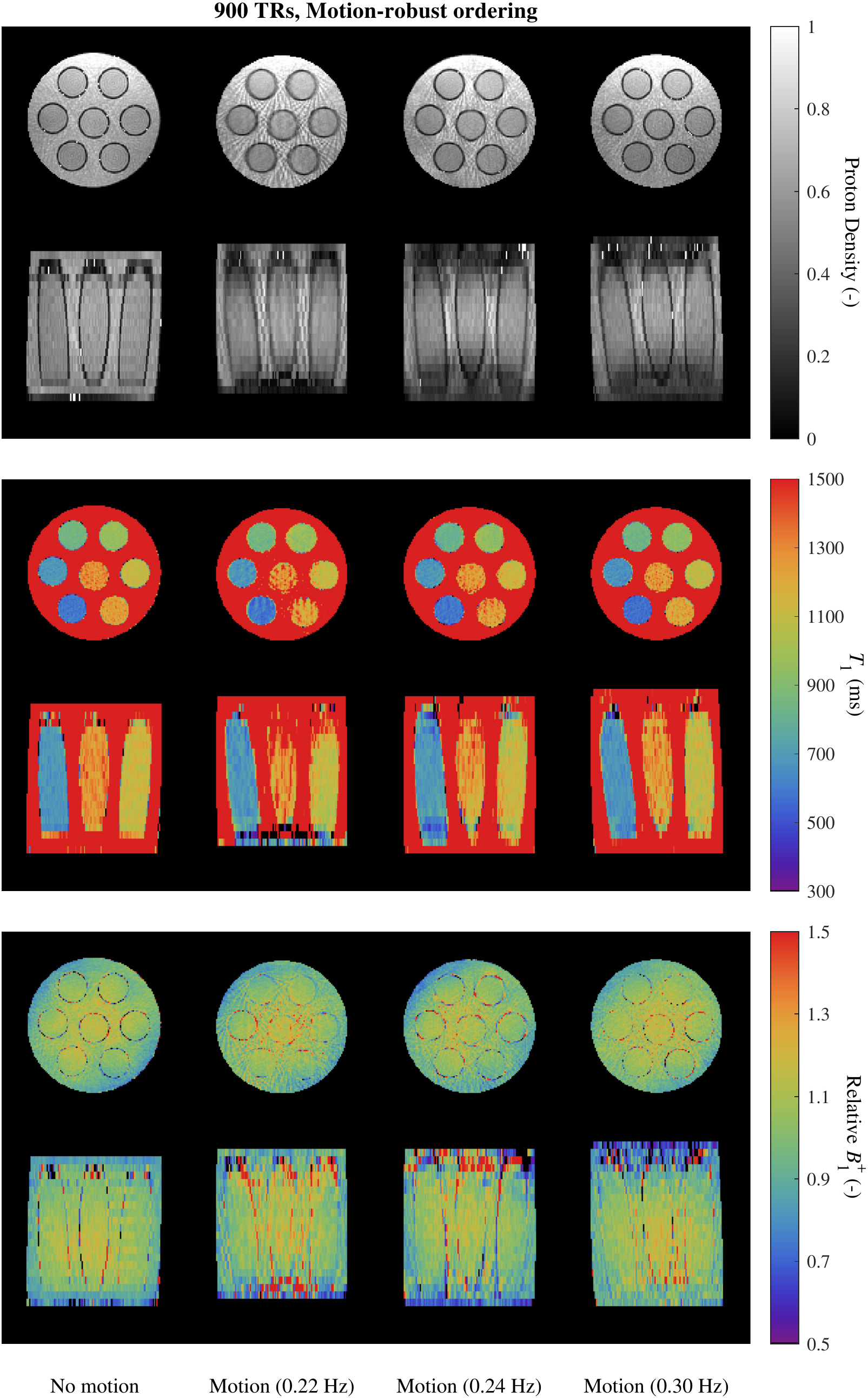}
    \caption{Quantitative Proton Density (top row), $T_1$ (middle row), and $B_1^+$ (bottom row) maps for the sequence with 900 TRs and the \textit{motion-robust ordering}. For each map, both a transverse (top) and a coronal (bottom) slice is shown.}
    \label{SIfig:900_mr}
\end{figure}

\begin{figure}[!t]
    \centering
    \includegraphics[width=12.7cm]{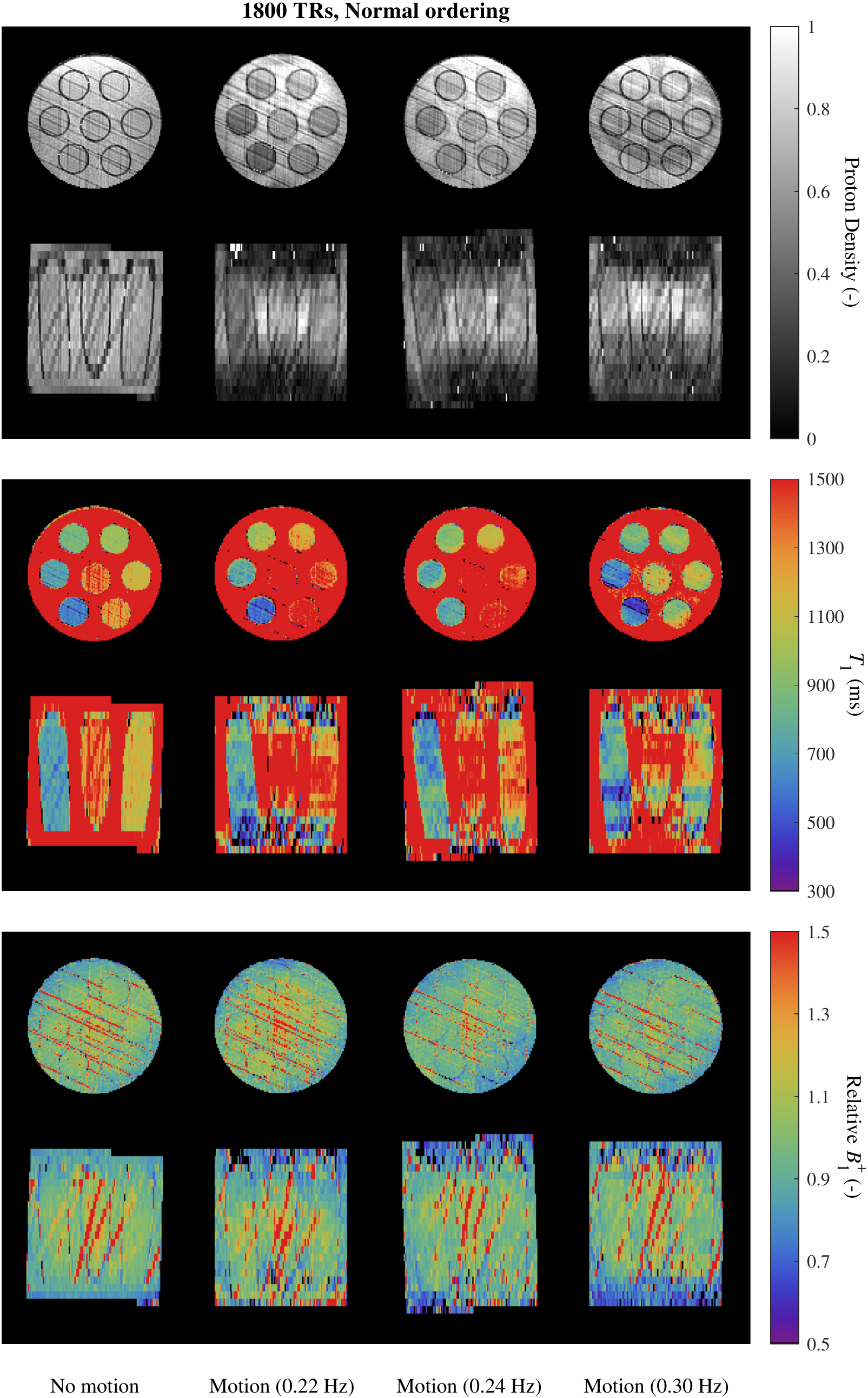}
    \caption{Quantitative Proton Density (top row), $T_1$ (middle row), and $B_1^+$ (bottom row) maps for the sequence with 1800 TRs and the \textit{normal ordering}. For each map, both a transverse (top) and a coronal (bottom) slice is shown.}
    \label{SIfig:1800_norm}
\end{figure}

\begin{figure}[!t]
    \centering
    \includegraphics[width=12.7cm]{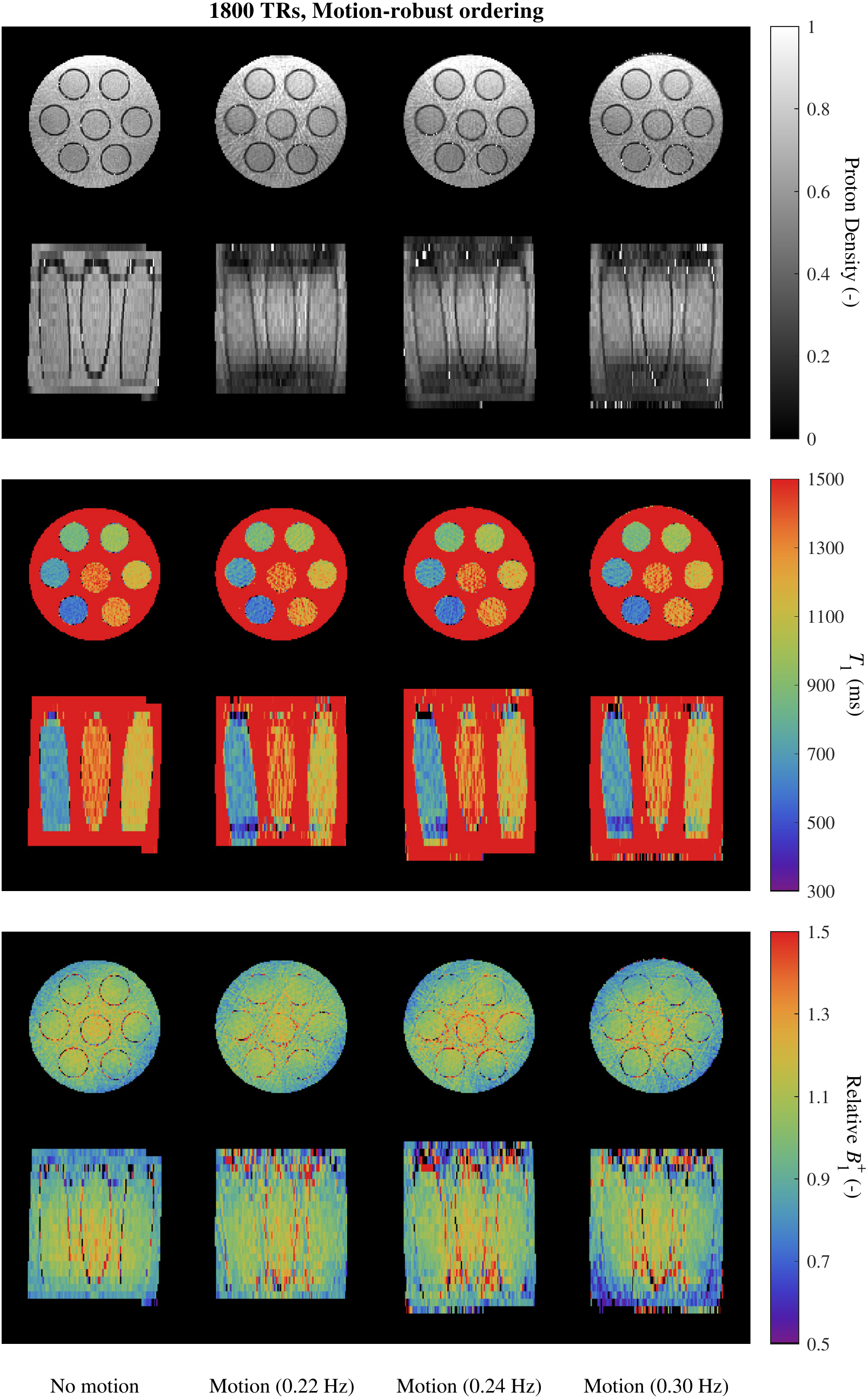}
    \caption{Quantitative Proton Density (top row), $T_1$ (middle row), and $B_1^+$ (bottom row) maps for the sequence with 1800 TRs and the \textit{motion-robust ordering}. For each map, both a transverse (top) and a coronal (bottom) slice is shown.}
    \label{SIfig:1800_mr}
\end{figure}

\begin{figure}[!t]
    \centering
    \includegraphics[width=10cm]{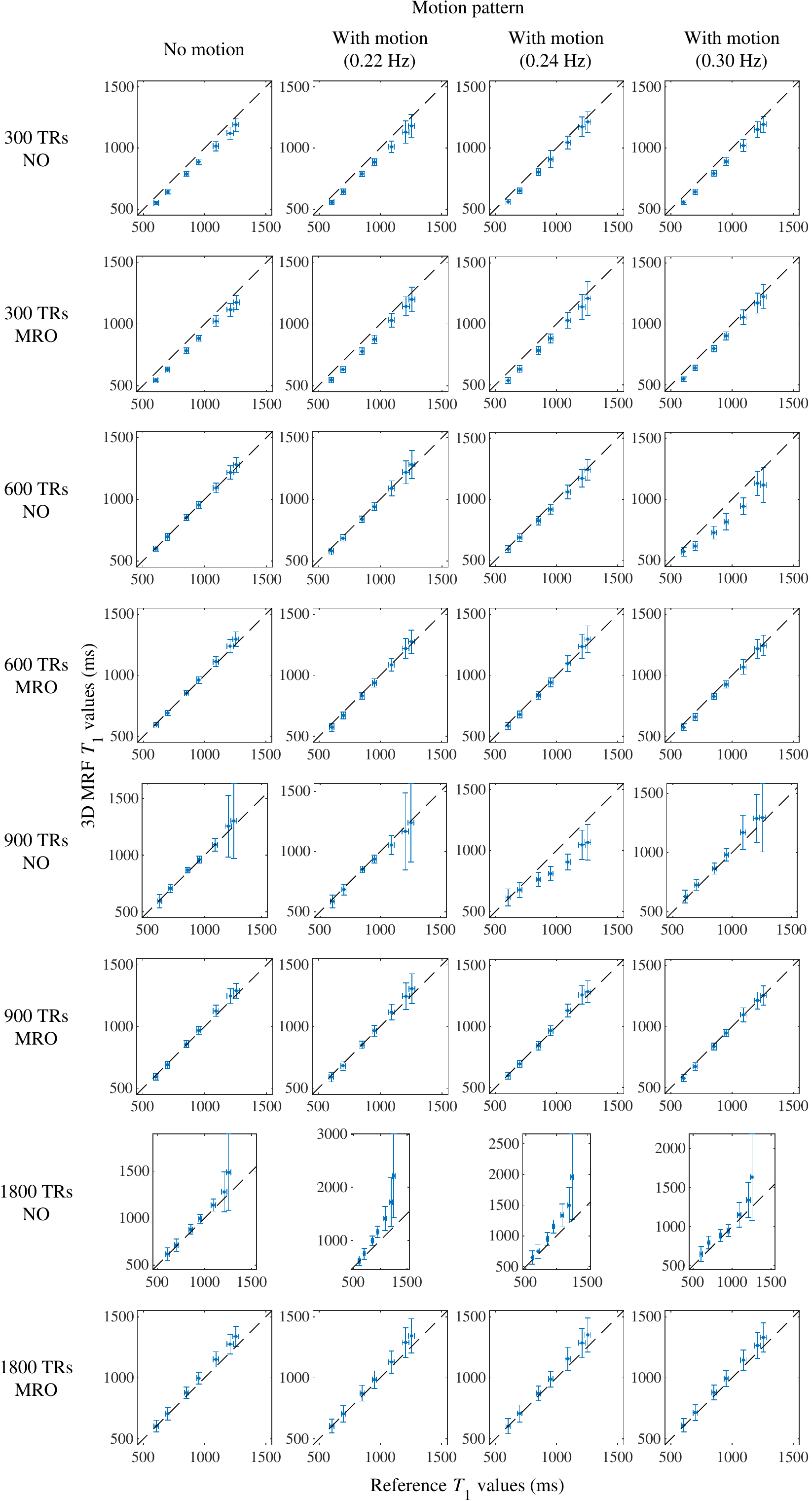}
    \captionof{figure}{Correlation plots between the estimated $T_1$ values (vertical axis) and the corresponding $T_1$ values from the reference scan (horizontal axis) for each map in Supporting Information Figures S\ref{SIfig:300_norm}-S\ref{SIfig:1800_mr}. Each point indicates the mean $T_1$ values of both scans, with the standard deviation depicted as error bars. The dashed line is the identity line. Note the increased deviation from the identity line for the normal ordering compared to the \textit{motion-robust ordering}.}
    \label{SIfig:correlations}
\end{figure}